\documentclass[runningheads]{llncs}
\usepackage{graphicx}
\usepackage[cmex10]{amsmath}
\usepackage{amssymb}
\usepackage{subcaption}
\usepackage[misc,geometry]{ifsym}

\usepackage{url}
\urldef{\mailsc}\path|haofu.liao@rochester.edu|

\graphicspath{ {./figures/data/} }
\begin{document}
\title{Generative Mask Pyramid Network for CT/CBCT Metal Artifact Reduction with \\ Joint Projection-Sinogram Correction}
\titlerunning{Generative Mask Pyramid Network for CT/CBCT Metal Artifact Reduction}

\author{Haofu Liao\textsuperscript{1(\Letter)}
\and Wei-An Lin\textsuperscript{2}
\and Zhimin Huo\textsuperscript{4}
\and Levon Vogelsang\textsuperscript{4}
\and \\ William J. Sehnert\textsuperscript{4}
\and S. Kevin Zhou\textsuperscript{3}
\and Jiebo Luo\textsuperscript{1}}

\institute{\textsuperscript{1} Department of Computer Science, University of Rochester\\ \mailsc\\
\textsuperscript{2} Department of ECE, University of Maryland, College Park\\
\textsuperscript{3} Institute of Computing Technology, Chinese Academy of Sciences\\
\textsuperscript{4} Carestream Health, Inc.}
\authorrunning{H. Liao et al}
%


%
\maketitle              

\begin{abstract}

A conventional approach to computed tomography (CT) or cone beam CT (CBCT) metal artifact reduction is to replace the X-ray projection data within the metal trace with synthesized data. However, existing projection or sinogram completion methods cannot always produce anatomically consistent information to fill the metal trace, and thus, when the metallic implant is large, significant secondary artifacts are often introduced. In this work, we propose to replace metal artifact affected regions with anatomically consistent content through \textit{joint projection-sinogram correction} as well as \textit{adversarial learning}. To handle the metallic implants of diverse shapes and large sizes, we also propose a novel \textit{mask pyramid network} that enforces the mask information across the network's encoding layers and a \textit{mask fusion loss} that reduces early saturation of adversarial training. Our experimental results show that the proposed projection-sinogram correction designs are effective and our method recovers information from the metal traces better than the state-of-the-art methods.

\end{abstract}
\section{Introduction}

Metal artifact is one of the most prominent artifacts which impede reliable computed tomography (CT) or cone beam CT (CBCT) image interpretation. It is commonly addressed in the \textit{sinogram domain} where the metal-affected regions in the sinograms are segmented and replaced with synthesized values so that metal-free CT images can be ideally reconstructed from the corrected sinograms. Early sinogram domain approaches fill the metal-affected regions by interpolation \cite{kalender1987reduction} or from prior images \cite{meyer2010normalized}. These methods can effectively reduce metal artifacts, but secondary artifacts are often introduced due to the loss of structural information in the corrected sinograms. Recent works propose to leverage deep neural networks (DNNs) to directly learn the sinogram correction. Park et al. \cite{park2017sinogram} applies U-Net \cite{ronneberger2015u} to correct metal-inserted sinogram, and Gjesteby et al. \cite{gjesteby2017projection} proposes to refine NMAR-corrected sinograms \cite{meyer2010normalized} using a convolutional neural network (CNN). Although better sinogram completions are achieved, the results are still subject to secondary artifacts due to the imperfect completion.

The development of DNNs in recent years also enables an \textit{image domain} approach that directly reduces metal artifacts or the related artifacts in CT/CBCT images. Specifically, the existing methods \cite{gjesteby2017reducing,zhang2018convolutional,xu2018deep,park2017,svar_gan} train image-to-image CNNs to transform artifact-affected CT images to artifact-free CT images. Gjesteby et al. \cite{gjesteby2017reducing} proposes to include the NMAR-corrected CT as the input with a two-stream CNN. Zhang et al. \cite{zhang2018convolutional} fuses beam hardening corrected and linear interpolated CT images for better correction. All the current image domain approaches use synthesized data to generate the metal-affected and metal-free image pairs for training. However, the synthesized data may not fully simulate the CT imaging under the clinical scenario making the image domain approaches less robust to clinical applications.

In this work, we propose a novel learning-based sinogram domain approach to metal artifact reduction (MAR). Unlike the existing image domain methods, the proposed method does not require synthesized metal artifact during training. Instead, we treat MAR as an image inpainting problem, i.e., we apply random metal traces to mask out artifact-free sinograms, and train a DNN to recover the data within the metal traces. Since metal-affected regions are viewed as missing, factors such as X-ray spectrum and the material of metal implants will not affect the generalizability of the proposed method. Unlike the existing learning-based sinogram domain approaches, our method delivers high-quality sinogram completion with three designs. \textit{First}, we propose a two-stage projection-sinogram\footnote{We denote the X-ray data that captured at the same view angle as a ``projection'' and a stack of projections corresponding to the same CT slice as a ``sinogram''.} completion scheme to achieve more contextually consistent correction results. \textit{Second}, we introduce adversarial learning into the projection-sinogram completion so that more structural and anatomically plausible information can be recovered from the metal regions. \textit{Third}, to make the learning more robust to the various shapes of metallic implants, we introduce a novel mask pyramid network (MPN) to distill the geometry information of different scales and a mask fusion loss to penalize early saturation. Our extensive experiments on both synthetic and clinical datasets demonstrate that the proposed method is indeed effective and perform better than the state-of-the-art MAR approaches. 
\section{Methodology}

\begin{figure}[t]
\begin{minipage}[b]{.48\textwidth}
  \includegraphics[width=\linewidth]{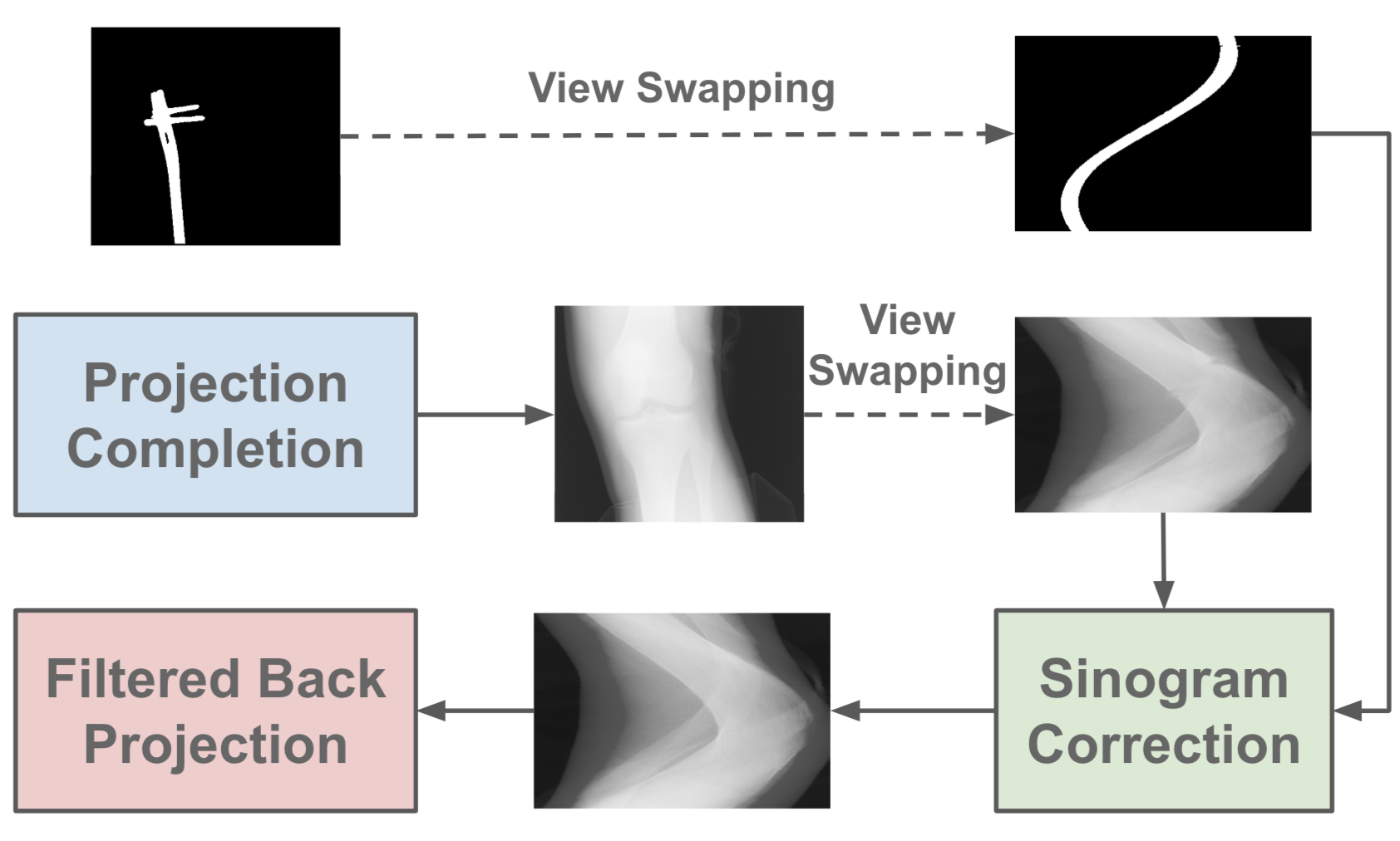}
  \centering
  \caption{Method overview.}
  \label{fig:overview}
\end{minipage}
\hfill
\begin{minipage}[b]{.48\textwidth}
  \includegraphics[width=\linewidth]{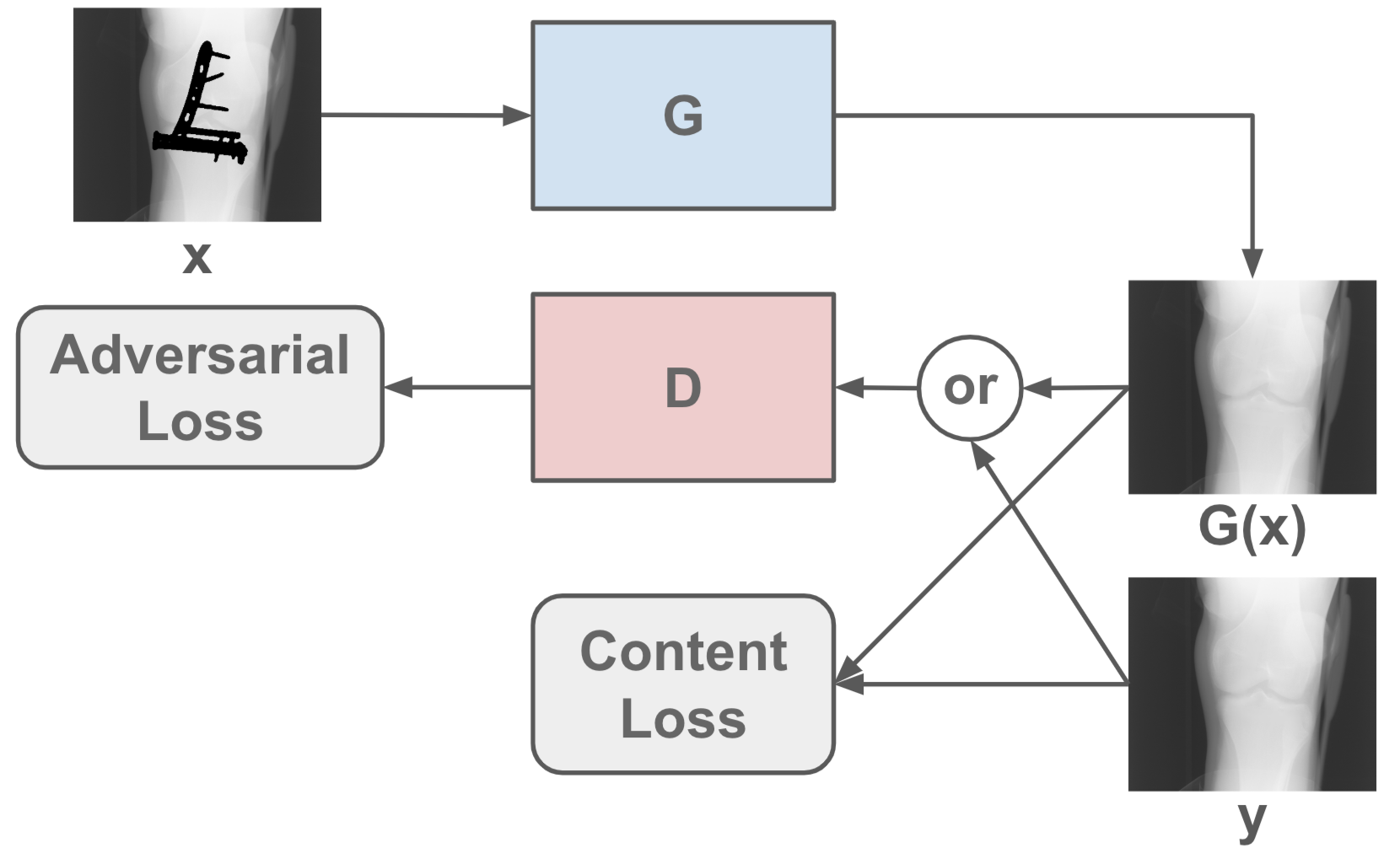}
  \centering
  \caption{The base framework.}
  \label{fig:architecture}
\end{minipage}
\end{figure}

An overview of the proposed method is shown in Fig. \ref{fig:overview}. Our method consists of two major modules: a projection completion module (blue) and a sinogram correction module (green). The projection completion module is an image-to-image translation model enhanced with a novel mask pyramid network. Given an input projection image and a pre-segmented metal mask, the projection completion module generates anatomically plausible and structurally consistent surrogates within the metal-affected regions. The sinogram correction module predicts a residual map to refine the projection-corrected sinograms. This joint projection-sinogram correction approach enforces inter-projection consistency and makes use of the context information between different viewing angles. Note that we perform projection completion first due to the observation that the projection images contain better structural information that facilitates the learning of an image inpainting model.

\subsubsection{Base Framework} \label{sec:basic_framework}

Inspired by recent advances in deep generative models \cite{pathak2016context,isola2017image}, we formulate the projection and sinogram correction problems under a generative image-to-image translation framework. The structure of the proposed model is illustrated in Fig. \ref{fig:architecture}. It consists of two individual networks: a generator $G$ and a discriminator $D$. The generator $G$ takes a metal-segmented projection $x$ as the input and generates a metal-free projection $G(x)$. The discriminator $D$ is a patch-based classifier that predicts if the metal-free projection $y$ or $G(x)$, is real or not. Similar to the PatchGAN \cite{isola2017image} design, $D$ is constructed as a CNN without fully-connected layers at the end to enable the patch-wise prediction. The detailed structures of $G$ and $D$ are presented in the supplementary material. $G$ and $D$ are trained adversarially with LSGAN \cite{mao2017least}, i.e.,

\begin{align}
  \label{eq:D1}
  \min_{D} \mathcal{L}_{\text{GAN}} &= \mathbb{E}_{y}[\lVert \mathbf{1} - D(y) \rVert^2] + \mathbb{E}_{x}[\lVert D(G(x)) \rVert^2], \\
  \label{eq:G1}
  \min_{G} \mathcal{L}_{\text{GAN}} &= \mathbb{E}_{x}[\lVert \mathbf{1} - D(G(x)) \rVert^2].
\end{align}
In addition, we also expect the generator output $G(x)$ to be close to its metal-free counterpart $y$. Therefore, we add a content loss $\mathcal{L}_c$ to ensure the pixel-wise consistency between $G(x)$ and $y$,
\begin{equation}
  \min_{G} \mathcal{L}_c = \mathbb{E}_{x,y}[\lVert G(x) - y \rVert_1].
\end{equation}

\begin{figure}[t]
\begin{minipage}[b]{.57\textwidth}
  \includegraphics[width=\linewidth]{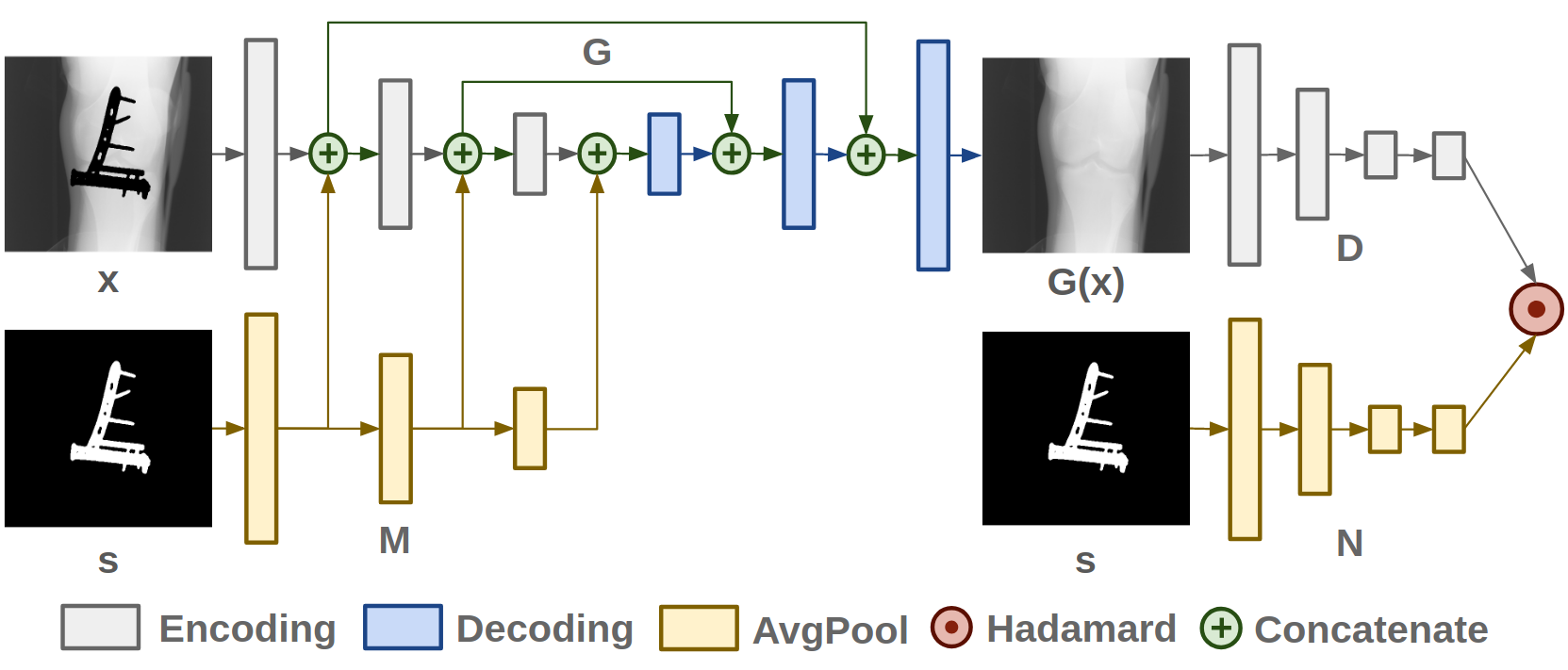}
  \centering
  \caption{Generator and discriminator.}
  \label{fig:generator_discriminator}
\end{minipage}
\begin{minipage}[b]{.38\textwidth}
  \includegraphics[width=\linewidth]{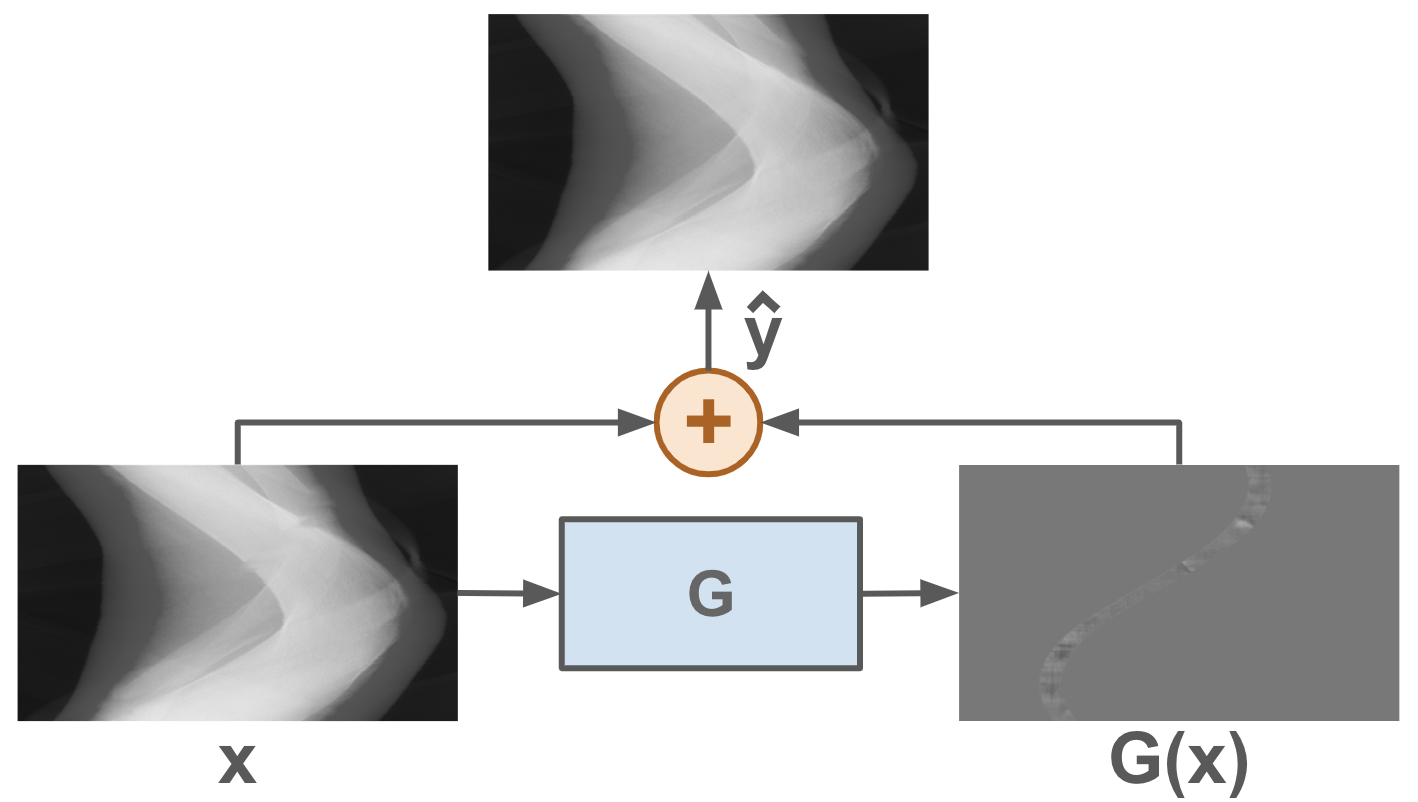}
  \centering
  \caption{Sinogram correction.}
  \label{fig:sinogram_correct}
\end{minipage}
\end{figure}


\subsubsection{Mask Pyramid Network}

Metallic implants have various shapes and sizes, such as metallic balls, bars, screws, wires, etc. When X-ray projections are acquired at different angles, the projected implants would exhibit complicated geometries. Hence, unlike typical image inpainting problems, where the shape of the mask is usually simple and fixed, projection completion is more challenging since the network has to learn how to fuse such diversified mask information of the metallic implants. Directly using metal-masked image as the input requires the metal mask information to be encoded by each layer and passed along to the later layers. For unseen masks, this encoding may not work very well and hence the mask information may be lost. To retain sufficient amount of mask information, we introduce a mask pyramid network (MPN) into the generator to feed the mask information into each layer explicitly.

The architecture of the generator $G$ with this design is illustrated in Fig. \ref{fig:generator_discriminator}. The MPN $M$ takes a metal mask $s$ as the input, and each block (in yellow) of $M$ is coupled with an encoding block (in grey) in $G$. Let $l^i_M$ denote the $i$th block of $M$ and $l^i_G$ denote the $i$th block of $G$. When $l^i_M$ and $l^i_G$ are coupled, the output of $l^i_M$ will be concatenated to the output of $l^i_G$. In this way, the mask information will then be used by $l^{i+1}_G$, and a recall of the mask is achieved. Each block $l^i_M$ of $M$ is implemented with an average pooling layer that has the same kernel, stride, and padding size as the convolutional layer in $l^{i}_G$. Hence, the metal mask output by $l^i_M$ not only has the same size as the feature maps from $l^{i}_G$, but also takes into account the receptive field of the convolution operation in $l^{i}_G$.

\subsubsection{Mask Fusion Loss} \label{sec:mask_focus}

In conventional image-to-image framework, the loss is usually computed on the entire image. On the one hand, this makes the generation less efficient, as a significant portion of the generator's computation will be spent on recovering the already known information. On the other hand, this also introduces early saturation during adversarial training, in which the generator stops improving in the masked regions, since the generator does not have information about the mask. We address this issue with two strategies. First, when computing the loss function, we only consider the content within the metal mask. That is, the content loss is rewritten as

\begin{equation}
  \min_{G} \mathcal{L}_c = \mathbb{E}_{x,y}[\lVert \hat{y} - y \rVert_1],
\end{equation}
where $\hat{y}= s \odot G(x) + (\mathbf{1} - s) \odot x$.

Second, we modulate the output score matrix from the discriminator by the metal mask $s$ so that the discriminator can selectively ignore the unmasked regions. As shown in Fig. \ref{fig:generator_discriminator}, we implement this design using another MPN $N$. But this time, we do not feed the intermediate outputs from $N$ to the coupled blocks in $D$, since the metal mask will, in the end, be applied to the loss. The adversarial part of the mask fusion loss is given as
\begin{equation} \label{eq:D2}
  \begin{split}
  \min_{D} \mathcal{L}_{\text{GAN}} = \mathbb{E}_{y}[\lVert N(s) \odot (\mathbf{1} - D(y)) \rVert^2] + \mathbb{E}_{x}[\lVert N(s) \odot D(\hat{y}) \rVert^2],
  \end{split}
\end{equation}
\begin{equation} \label{eq:G2}
  \min_{G} \mathcal{L}_{\text{GAN}} = \mathbb{E}_{x}[\lVert N(s) \odot (\mathbf{1} - D(\hat{y})) \rVert^2],
\end{equation}
and the total mask fusion loss can be written as
\begin{equation} \label{eq:total}
\mathcal{L} = \mathcal{L}_{\text{GAN}} + \lambda \mathcal{L}_c,
\end{equation}
where $\lambda$ balances the importance between $\mathcal{L}_{\text{GAN}}$ and $\mathcal{L}_c$.

\subsubsection{Sinogram Correction with Residual Map Learning} \label{sec:sinogram_correct}

Although the proposed projection completion framework in previous sections can produce an anatomically plausible result, it only considers the contextual information within a projection. Observing that a stack of consecutive projections form a set of sinograms. We use a simple yet effective model to enforce the inter-projection consistency by having the completion results look like sinograms.

Let $x$ denote a sinogram formed from previous projection completion step. A generator, as shown in Fig. \ref{fig:sinogram_correct}, predicts a residual map $G(x)$ which is then added to $x$ to correct the projection completion results. Here, we use the same generator structure as the one introduced in Fig. \ref{fig:generator_discriminator}. For the objective function, we apply the same one as used in Eq. \ref{eq:total}, except that we have $\hat{y} = s \odot (G(x) + x)  + (\mathbf{1} - s) \odot x$.

\section{Experimental Evaluations}

\subsubsection{Implementation Details and Baselines}

We implement the proposed model using PyTorch and train the model with the Adam optimization method. For the hyper-parameters, we set learning rate $= 5e^{-4}$, $\beta_1=0.5$, $\lambda=100$, and batch size $=16$. We compare our projection completion (PC) model and joint projection-sinogram correction (PC+SC) model with the following baseline MAR approaches: 1) LI, sinogram correction by linear interpolation \cite{kalender1987reduction}; 2) BHC, beam hardening correction for MAR \cite{verburg2012ct}; 3) NMAR, a state-of-the-art MAR model \cite{meyer2010normalized} that produces a prior CT image to correct metal artifacts; and 4) CNNMAR, the state-of-the-art deep learning based method \cite{zhang2018convolutional} that uses a CNN to output the prior image for MAR. 

\subsubsection{Datasets and Simulation Details}

For the synthesized dataset, we use the images collected from a CBCT scanner that is dedicated for lower extremities.
The size of the CBCT projections is $448\times512$ and the projections contain no metal objects. We randomly apply masks to the projections to obtain masked and unmasked projection pairs. In total, there are 27 CBCT scans, each with 600 projections. Projections from 24 of the CBCT scans are used for training, and the rest are held out for testing.


Two types of object masks are collected for the experiments: metal masks and blob masks. For the metal masks, we collect 3D binary metal implant volumes from clinical records and forward project them to obtain 2D metal projection masks. In total, we obtain 18,000 projection masks from 30 binary metal implant volumes. During training, we simulate the metal implants insertion process by randomly placing metal segmentation masks on the metal-free projections. For the blob masks, we adopt the method from \cite{pathak2016context} by drawing randomly shaped blobs on the image. Results for projection and sinogram completion with the metal and blob masks are provided in the supplementary material.



For a fair comparison, we adopt the same procedures as in \cite{zhang2018convolutional} to synthesize metal-affected CBCT volumes. We assume a 120 kVp X-ray source with $2 \times 10^7$ photons. The distance from the X-ray source to the rotation center is set to 59.5cm, and 416 projection views are uniformly spaced between 0-360 degrees. The size of the reconstructed volume is $448\times448\times448$. During simulation, we set the material to iron for all the metal masks. Note that since the metal masks are from clinical records, the geometries and intensities of the metal artifacts are extremely diverse, which makes MAR highly challenging.

For the clinical dataset, we use the vertebrae localization and identification dataset from Spineweb\footnote{spineweb.digitalimaginggroup.ca}. 
We first define regions with HU values greater than 2,500 as metal regions. Then, we select images with the largest-connected metal region greater than 400 pixels as metal-affected images and images with the largest HU value smaller than 2,000 as metal-free images. The metal masks for the projections and sinograms are obtained by forward projecting the metal regions in the CT image domain. The training for this dataset is performed on the metal-free images with metal masks obtained from the metal-affected images.

\begin{figure}[t]
    \centering
    \begin{subfigure}[b]{.48\linewidth}
        \includegraphics[width=\textwidth]{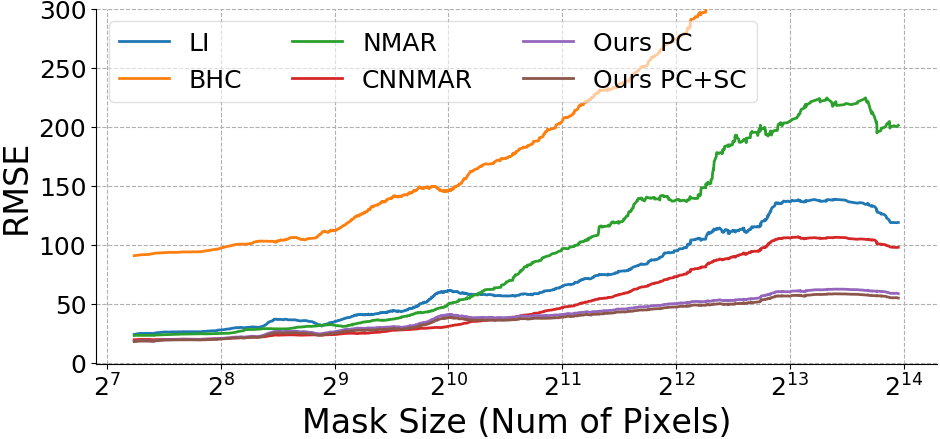}
        \caption{RMSE}
    \end{subfigure}
    \begin{subfigure}[b]{.48\linewidth}
        \includegraphics[width=\textwidth]{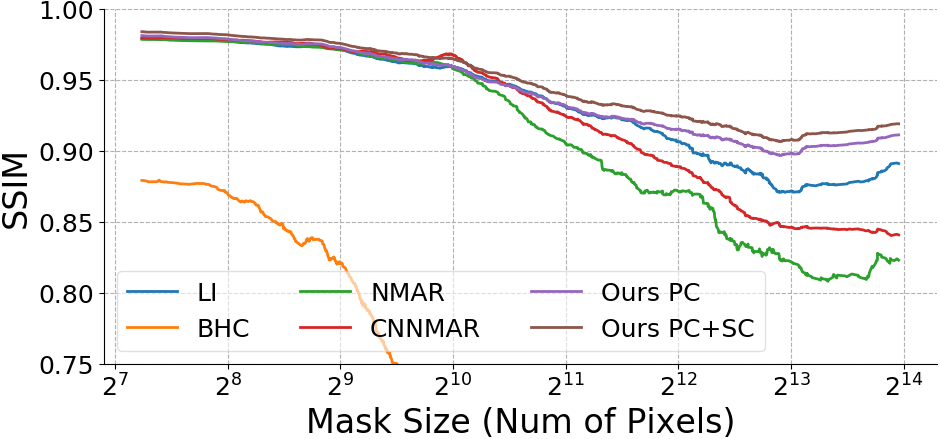}
        \caption{SSIM}
    \end{subfigure}
    \caption{Quantitative MAR results of different models with respect to different mask sizes. For RMSE/SSIM, the lower/higher values are better.}
    \label{fig:mar_metrics}
\end{figure}
\begin{figure}[t]
  \centering
  \begin{subfigure}[b]{.24\linewidth}
      \captionsetup{justification=centering}
      \includegraphics[width=\textwidth]{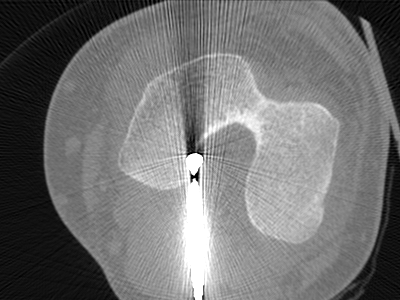}
      \caption{Input: 286/0.73}
  \end{subfigure}
  \begin{subfigure}[b]{.24\linewidth}
      \captionsetup{justification=centering}
      \includegraphics[width=\textwidth]{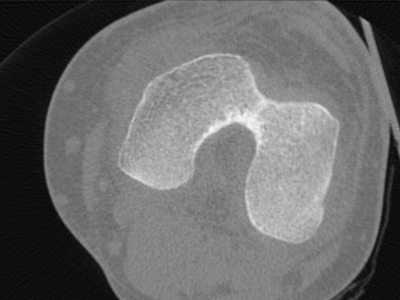}
      \caption{Ground Truth}
  \end{subfigure}
  \begin{subfigure}[b]{.24\linewidth}
      \captionsetup{justification=centering}
      \includegraphics[width=\textwidth]{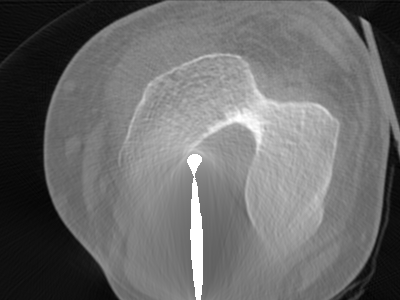}
      \caption{LI: 79/0.93}
  \end{subfigure}
  \begin{subfigure}[b]{.24\linewidth}
      \captionsetup{justification=centering}
      \includegraphics[width=\textwidth]{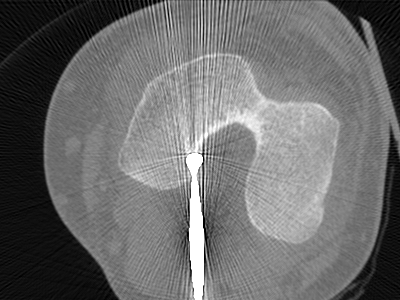}
      \caption{BHC: 226/0.69}
  \end{subfigure}
  \begin{subfigure}[b]{.24\linewidth}
      \captionsetup{justification=centering}
      \includegraphics[width=\textwidth]{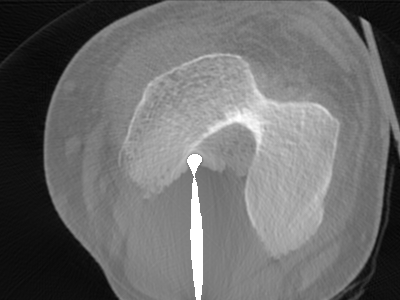}
      \caption{NMAR: 57/0.93}
  \end{subfigure}
  \begin{subfigure}[b]{.24\linewidth}
      \captionsetup{justification=centering}
      \includegraphics[width=\textwidth]{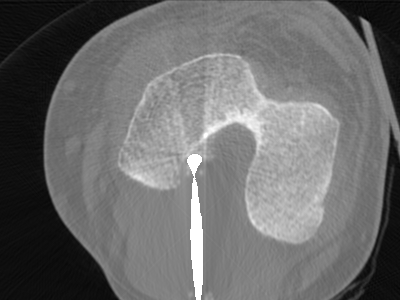}
      \caption{\tiny{CNNMAR} \normalsize{:41/0.92}}
  \end{subfigure}
  \begin{subfigure}[b]{.24\linewidth}
      \captionsetup{justification=centering}
      \includegraphics[width=\textwidth]{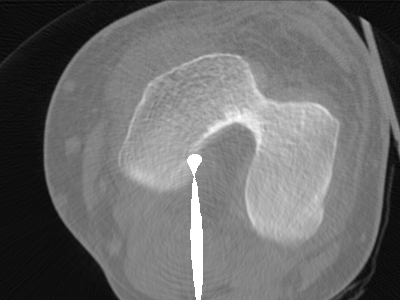}
      \caption{PC: 30/0.93}
  \end{subfigure}
  \begin{subfigure}[b]{.24\linewidth}
      \captionsetup{justification=centering}
      \includegraphics[width=\textwidth]{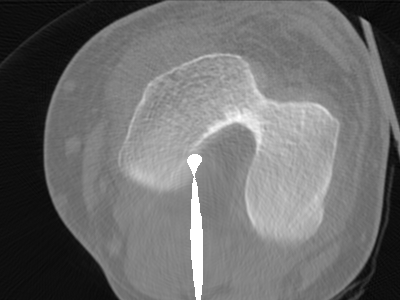}
      \caption{PC+SC: 29/0.94}
  \end{subfigure}
  \caption{MAR results on images with synthesized metal artifacts. Metallic implants are replaced with constant values (white) after MAR. The reported numbers are RMSE (HU) / SSIM.}
  \label{fig:sample_results}
\end{figure}

\subsubsection{Quantitative Comparisons}  We use two metrics: the rooted mean square error (RMSE) and structural similarity index (SSIM) for quantitative evaluations. We conduct a thorough study by evaluating RMSE and SSIM for a wide range of mask sizes. The results are summarized in Fig. \ref{fig:mar_metrics}. We observe that the proposed method achieves superior performance over the other methods. For example, the RMSE error of the second-best method CNNMAR \cite{zhang2018convolutional} almost doubles that of the proposed method when the implant size is large. In addition, by further refining in the sinogram domain, improved performance can be achieved especially in terms of the SSIM metric. From Fig. \ref{fig:mar_metrics}, we also perceive that methods which require tissue segmentation (e.g. NMAR and CNNMAR) perform well when the metallic object is smaller than $1000$ pixels. However, when the size of the metallic implants becomes larger, these methods deteriorate significantly due to erroneous segmentation. The proposed joint correction approach, which does not rely on tissue segmentation, exhibits less degradation.

\subsubsection{Qualitative Comparisons}

\begin{figure}[t]
\centering
\includegraphics[width=0.8\textwidth]{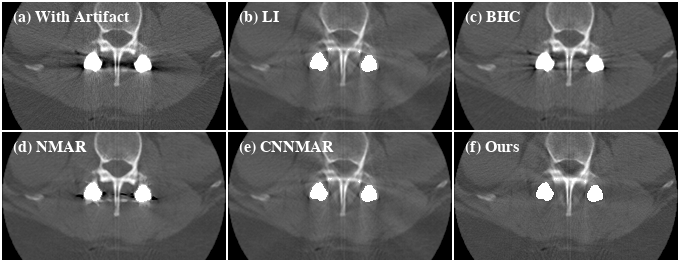}
\caption{MAR results on clinical images. Metallic implants are replaced with constant values (white) after MAR.}
\label{fig:clinical_results}
\end{figure}

Fig. \ref{fig:sample_results} shows MAR results on synthesized metal-affected images. It is clear that the proposed method successfully restores streaking artifacts caused by metallic implants. Unlike other approaches that generates erroneous surrogates, our method fills in contextually consistent values through generative modeling and joint correction. For the results with clinical data (Fig \ref{fig:clinical_results}), we also observe that our method produces qualitatively better results. BHC and NMAR cannot totally reduce the metal artifacts. LI and CNNMAR can recover most of the metal-affected regions. However, they also produce secondary artifacts. We notice a performance degradation for CNNMAR on the clinical data compare to the synthesized data, which demonstrates that image domain approaches relying on synthesizing metal artifact have worse generalizability.

\section{Conclusion}

We present a novel MAR approach based on a generative adversarial framework with joint projection-sinogram correction and mask pyramid network. From the experimental evaluations, we show that existing MAR methods does not effectively reduce metal artifact. By contrast, the proposed approach leverages the extra contextual information from sinogram and achieves a superior performance over other MAR methods in both the synthesized and clinical datasets.

\noindent \textbf{Acknowledgement.} This work was supported in part by NSF award \#1722847, the Morris K. Udall Center of Excellence in Parkinson's Disease Research by NIH, and the corporate sponsor Carestream.

%
%

\bibliographystyle{splncs04}
\bibliography{references}

\newpage

\noindent \vspace{.1em}
\begin{center}
\textbf{\LARGE{Supplementary Material}}\vspace{3em}
\end{center}

\renewcommand\thesection{\Alph{section}}
\setcounter{section}{0}

\begin{figure}[h]
    \centering
    \begin{subfigure}[b]{.83\linewidth}
        \includegraphics[width=\textwidth]{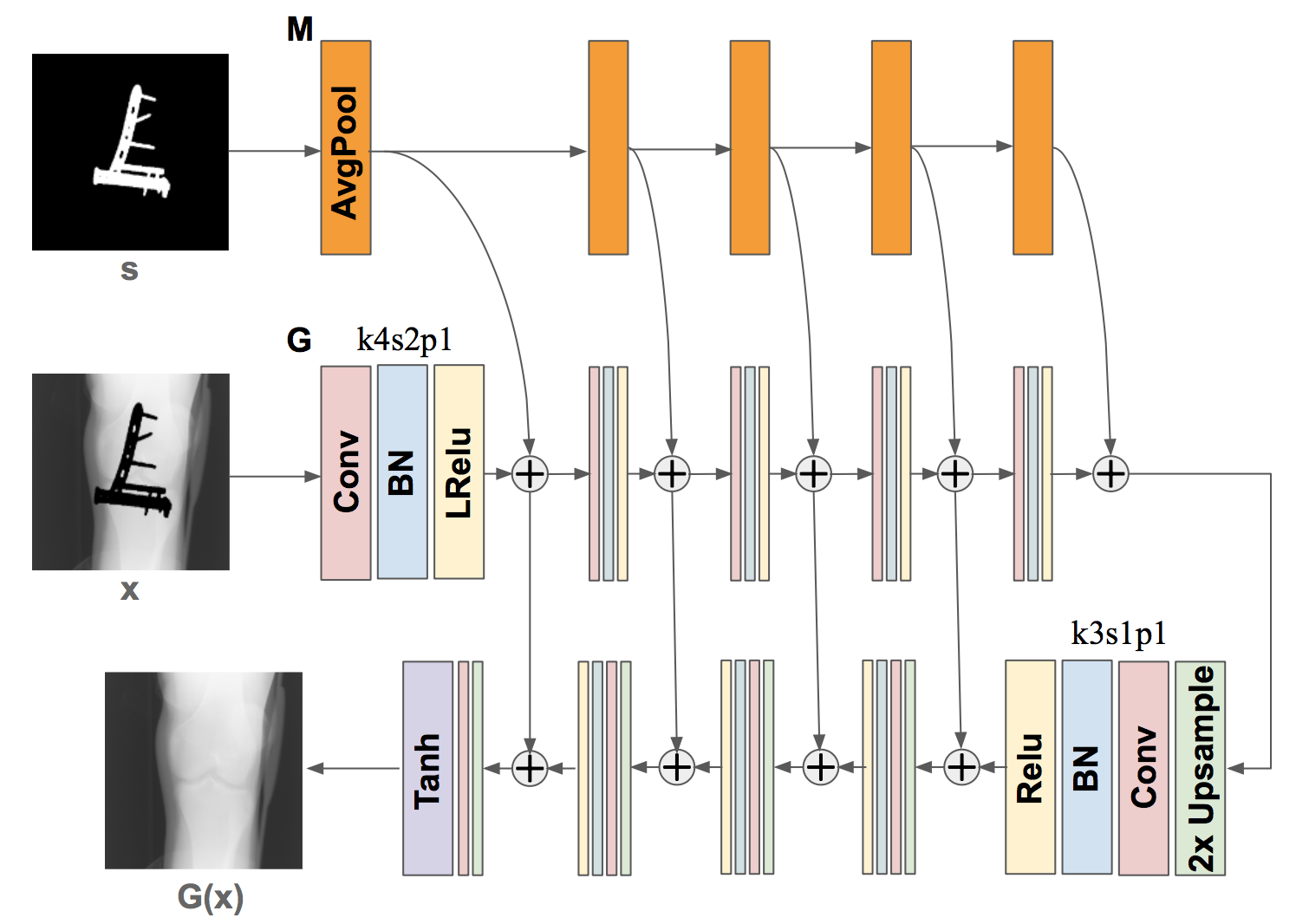}
        \caption{Generator.}
    \end{subfigure}
    \\
    \begin{subfigure}[b]{.83\linewidth}
        \includegraphics[width=\textwidth]{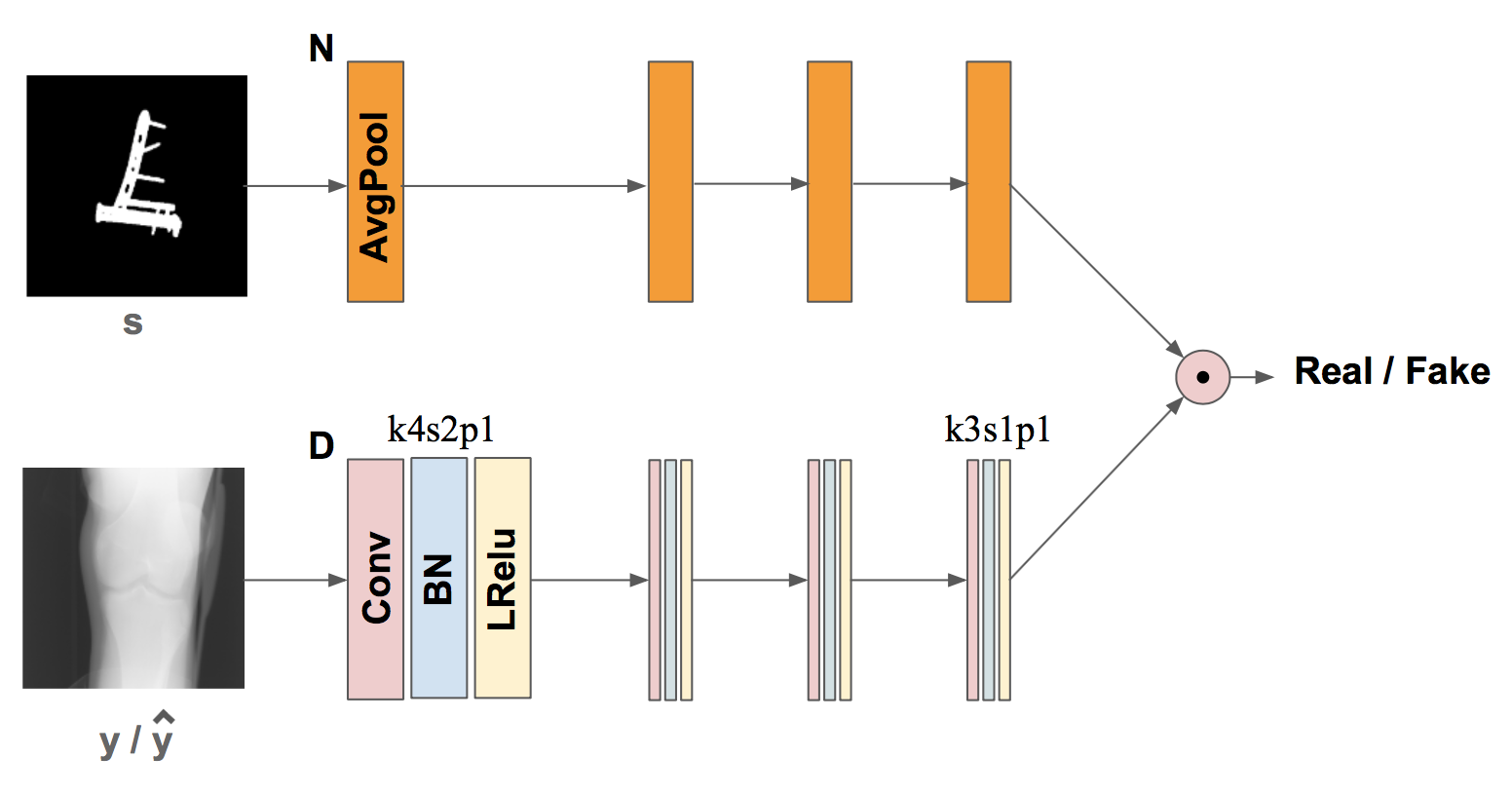}
        \caption{Discriminator.}
    \end{subfigure}
    \caption{Detailed network architecture. k: kernel, s: stride, and p: padding sizes.}
\end{figure}
\vspace{-0.5cm}
\begin{figure}[h]
    \centering
    \begin{subfigure}[b]{.28\linewidth}
      \includegraphics[width=\textwidth]{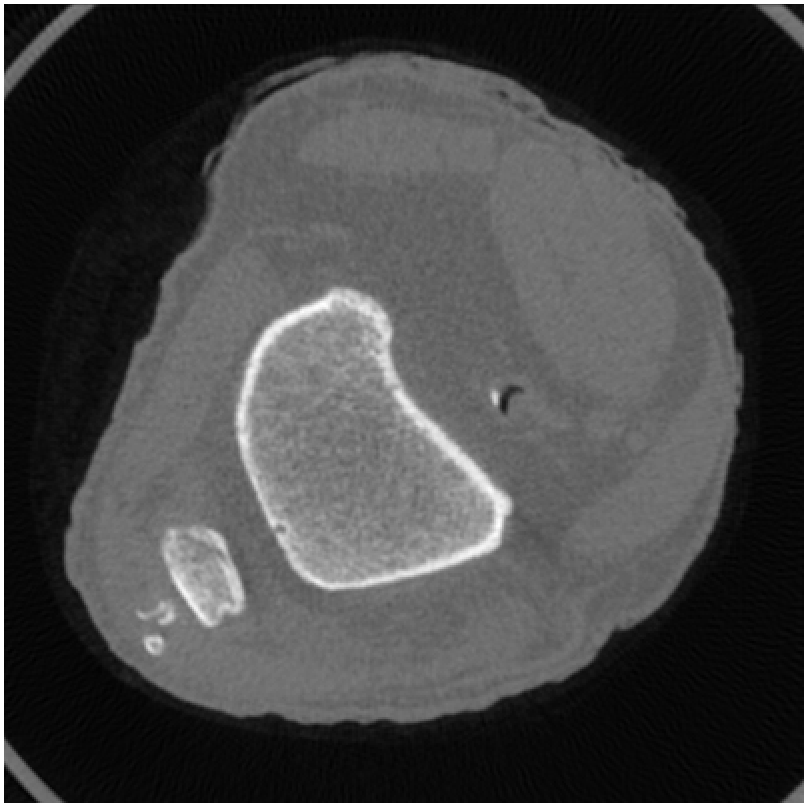}
  \end{subfigure}
  \begin{subfigure}[b]{.28\linewidth}
      \includegraphics[width=\textwidth]{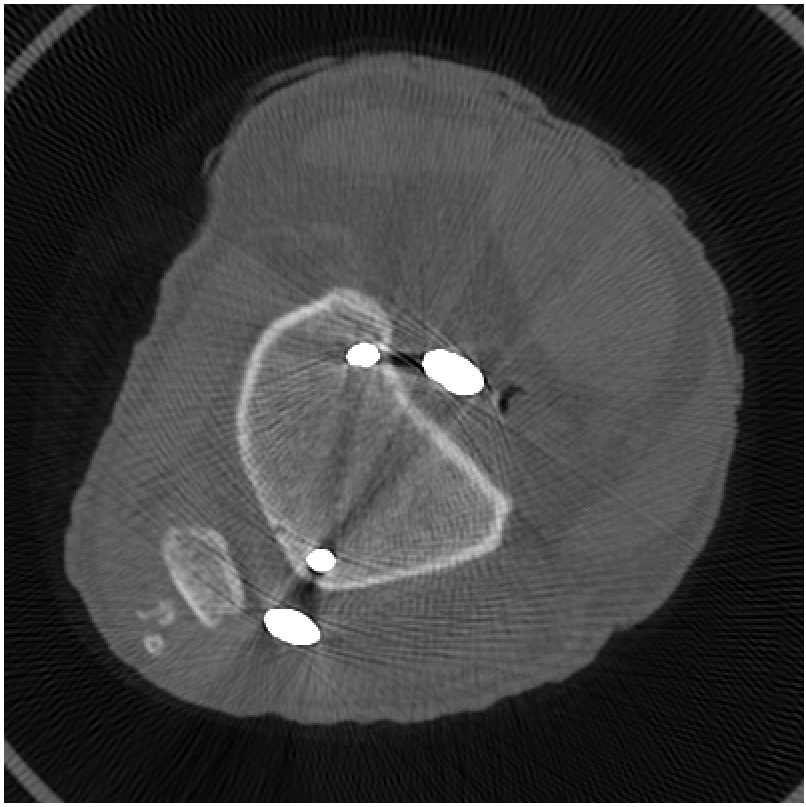}
  \end{subfigure}
    \begin{subfigure}[b]{.28\linewidth}
        \includegraphics[width=\textwidth]{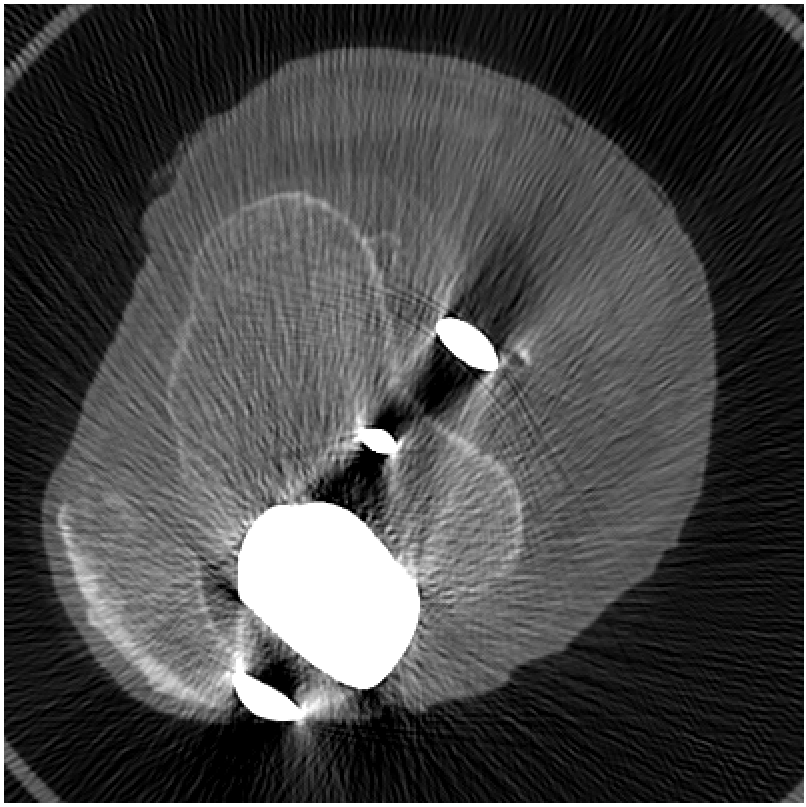}
    \end{subfigure}
    \caption{Samples from synthesized CBCT image dataset with (a) no, (b) mild, and (c) severe artifact.}
    \label{fig:simulated_images}
\end{figure}
\vspace{-1cm}
\begin{figure}[h]
  \centering
  \begin{subfigure}[b]{.3\linewidth}
    \includegraphics[width=\textwidth]{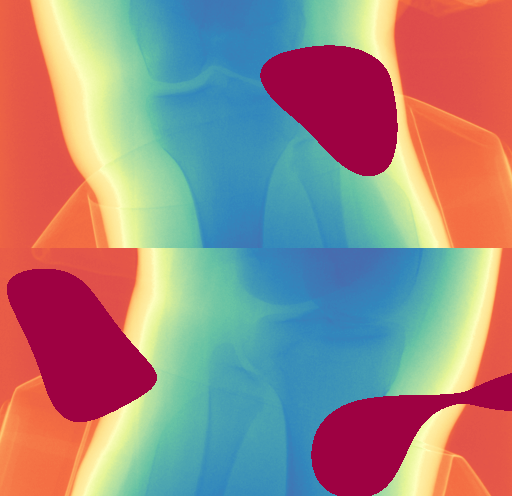}
    \caption{Masked}
\end{subfigure}
\begin{subfigure}[b]{.3\linewidth}
    \includegraphics[width=\textwidth]{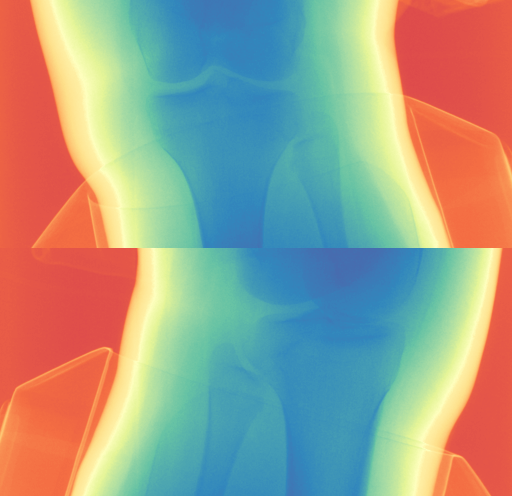}
    \caption{Unmasked}
\end{subfigure}
  \begin{subfigure}[b]{.3\linewidth}
      \includegraphics[width=\textwidth]{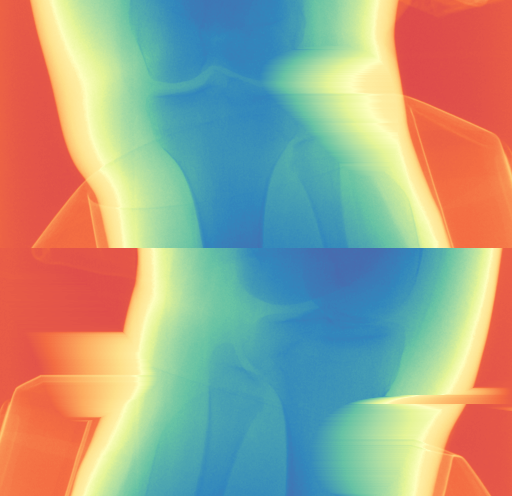}
      \caption{LI}
  \end{subfigure}
  \\
  \begin{subfigure}[b]{.3\linewidth}
      \includegraphics[width=\textwidth]{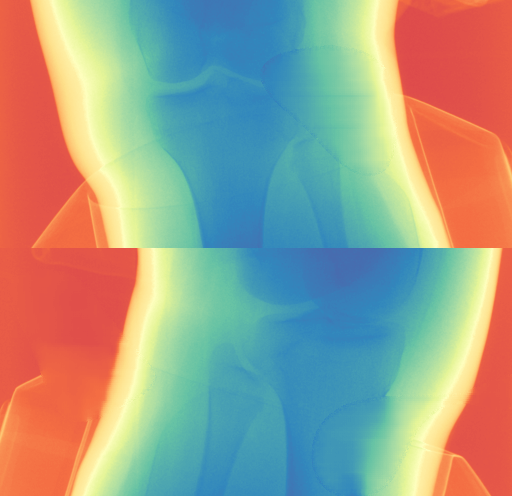}
      \caption{BF.}
  \end{subfigure}
  \begin{subfigure}[b]{.3\linewidth}
    \includegraphics[width=\textwidth]{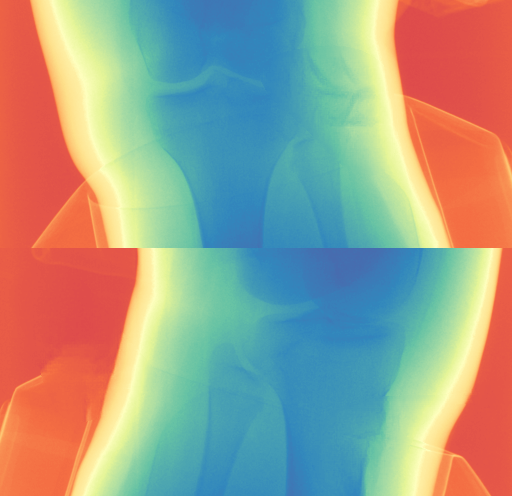}
    \caption{With MFL.}
\end{subfigure}
  \caption{Visual comparison of models completing blob-masked X-ray projections. 
  (c) LI: linear interpolation (d) BF: base framework. (e) MFL: base framework with mask fusion loss. A colormap is applied for improved details of the bone region.}
  \label{fig:compare_mask_focused}
\end{figure}

\begin{figure}[h]
  \centering
  \begin{subfigure}[b]{.3\linewidth}
    \includegraphics[width=\textwidth]{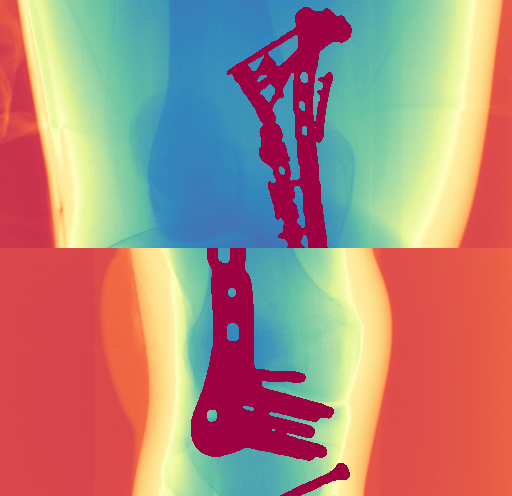}
    \caption{Masked}
\end{subfigure}
\begin{subfigure}[b]{.3\linewidth}
    \includegraphics[width=\textwidth]{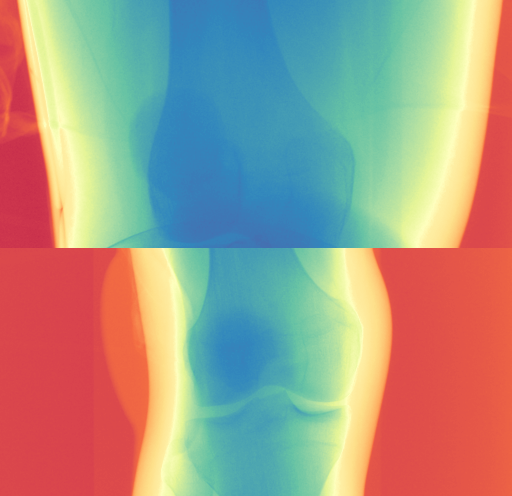}
    \caption{Unmasked}
\end{subfigure}
  \begin{subfigure}[b]{.3\linewidth}
      \includegraphics[width=\textwidth]{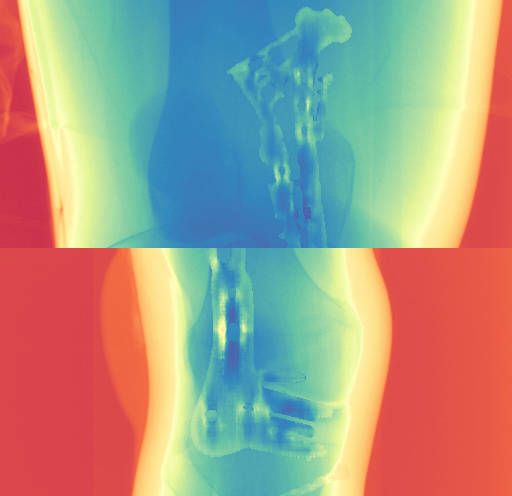}
      \caption{MFL-blob}
  \end{subfigure}
  \\
  \begin{subfigure}[b]{.3\linewidth}
      \includegraphics[width=\textwidth]{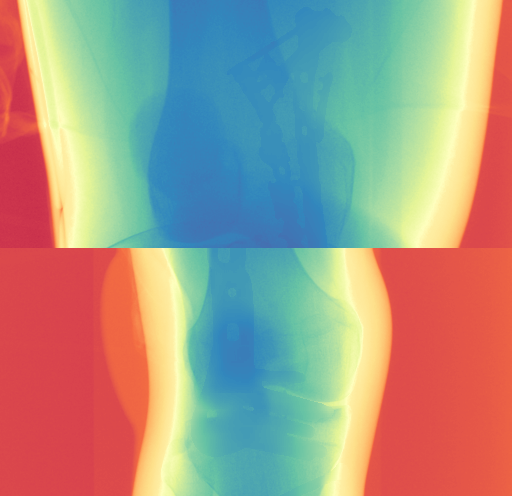}
      \caption{MFL-metal}
  \end{subfigure}
  \begin{subfigure}[b]{.3\linewidth}
    \includegraphics[width=\textwidth]{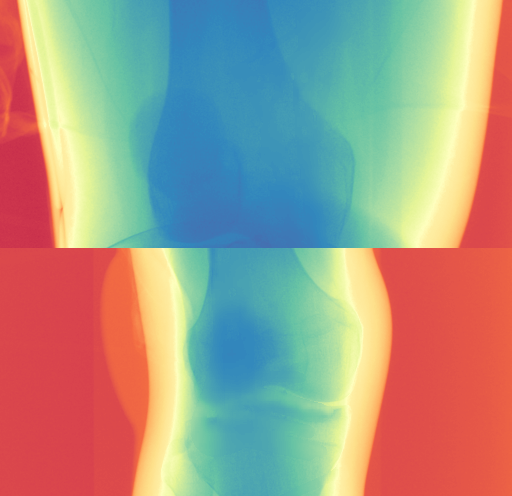}
    \caption{PC}
\end{subfigure}
  \caption{Visual comparison of models completing metal-masked X-ray projections. (c) MFL trained using blob-masked projections. (d) MFL trained using metal-masked projections. (e) MFL with mask pyramid network trained using metal-masked projections. A colormap is applied for improved details of the bone region.}
  \label{fig:compare_mask_recalled}
\end{figure}

\begin{figure}[t]
    \centering
    \begin{subfigure}[b]{.24\linewidth}
        \includegraphics[width=\textwidth]{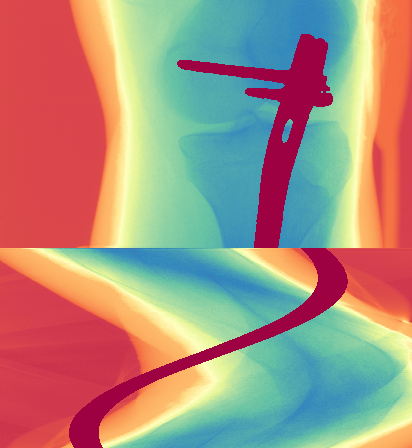}
        \caption{Masked}
    \end{subfigure}
    \begin{subfigure}[b]{.24\linewidth}
        \includegraphics[width=\textwidth]{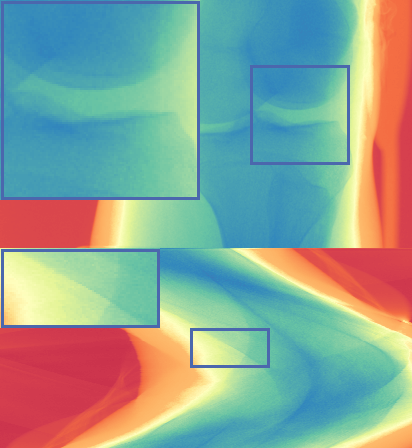}
        \caption{Unmasked}
    \end{subfigure}
    \begin{subfigure}[b]{.24\linewidth}
        \includegraphics[width=\textwidth]{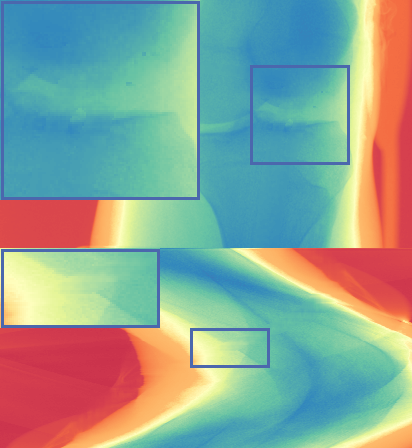}
        \caption{PC}
    \end{subfigure}
    \begin{subfigure}[b]{.24\linewidth}
        \includegraphics[width=\textwidth]{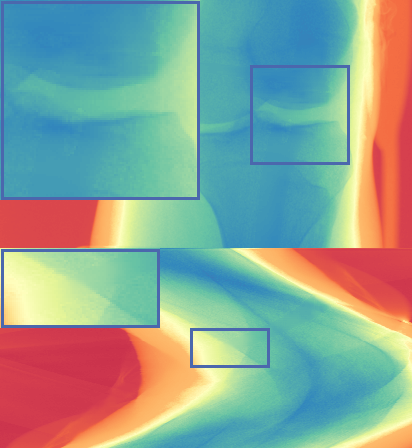}
        \caption{PC+SC}
    \end{subfigure}
    \caption{Projection (top) and sinogram (bottom) completion results with and without SC. A colormap is applied for improved details of the bone region.}
    \label{fig:sinogram_refine}
\end{figure}

\end{document}